\documentclass[conference]{IEEEtran}
\IEEEoverridecommandlockouts
\usepackage{cite}
\usepackage{amsmath,amssymb,amsfonts}
\usepackage{mathtools}
\usepackage[ruled,vlined]{algorithm2e}
\usepackage{graphicx}
\usepackage{textcomp}
\usepackage[table]{xcolor}
\usepackage[bookmarks=false]{hyperref}
\usepackage[nolist]{acronym}
\usepackage{tabularx}
\usepackage{booktabs}
\usepackage{soul}
\usepackage{makecell}

\begin{acronym}
    \acro{AWGN}{additive white Gaussian noise}
    \acro{ISAC}{integrated sensing and communication}
    \acro{DFRC}{dual-function radar communication}
    \acro{AoA}{angle of arrival}
    \acro{PR}{passive radio}
    \acrodefplural{AoA}[AoAs]{angles of arrival}
    \acro{ToA}{time of arrival}
    \acrodefplural{ToA}[ToAs]{times of arrival}
    \acro{AoD}{angle of departure}
    \acrodefplural{AoD}[AoDs]{angles of departure}
	\acro{LoS}{line-of-sight}
	\acro{NLoS}{non-line-of-sight}
	\acro{ULA}{uniform linear array}
	\acro{OFDM}{orthogonal frequency-division multiplexing}
	\acro{CRB}{Cram\'er-Rao bound}
	\acro{i.i.d}{independently and identically distributed}
	\acro{DFT}{discrete Fourier transform}
	\acro{SISO}{single-input, single-output}
    \acro{MIMO}{multiple-input, multiple-output}
	\acro{MSE}{mean squared error}
    \acro{CM}{channel modeling}
	\acro{SNR}{signal-to-noise ratio}
    \acro{SINR}{signal-to-interference-and-noise ratio}
	\acro{BGG}{bistatic geometric gain}
	\acro{ESNR}{estimation signal-to-noise ratio}
	\acro{DMC}{dense multipath components}
	\acro{SC}{specular components}
	\acro{RCS}{radar cross section}
	\acro{TF}{transfer function}
	\acro{PDP}{power-delay profile}
	\acro{PAP}{power-angular profile}
	\acro{WB}{wideband}
	\acro{UWB}{ultra-wideband}
	\acro{SV}{Saleh-Valenzuela}
	\acro{AS}{angular spread}
	\acro{DS}{delay spread}
	\acro{FCF}{frequency correlation function}
	\acro{VMD}{Von-Mises distribution}
	\acro{FIM}{Fisher information matrix}
	\acro{LM}{Levenberg-Marquardt}
	\acro{MRC}{maximal ratio combining}
	\acro{RMSE}{root mean squared error}
	\acro{Tx}{transmitter}
	\acro{Rx}{receiver}
	\acro{THz}{terahertz}
	\acro{UE}{user equipment}
	\acro{BS}{base station}
	\acro{PD}{probability of detection}
	\acro{PFA}{probability of false alarm}
	\acro{CSI}{channel state information}
\end{acronym}

\definecolor{gray10pct}{gray}{0.9}

\DeclareMathOperator{\BW}{BW}

\DeclareMathOperator{\toep}{toep}

\DeclareMathOperator{\CRB}{CRB}

\DeclareMathOperator{\cat}{cat}
\DeclareMathOperator{\NF}{NF}

\DeclareMathOperator*{\blockdiag}{blkdiag}

\newcommand{\norm}[1]{\left\lVert#1\right\rVert}
\newcommand{\magn}[1]{\left\lvert#1\right\rvert}
\usepackage{xcolor}

\def\BibTeX{{\rm B\kern-.05em{\sc i\kern-.025em b}\kern-.08em
    T\kern-.1667em\lower.7ex\hbox{E}\kern-.125emX}}
\begin{document}

\title{Impact of Background Dense Multipath Components on Multi-Band Fusion ISAC Systems}

\author{
\IEEEauthorblockN{
Dexin Wang\IEEEauthorrefmark{1}\IEEEauthorrefmark{2}, 
Roberto Bomfin\IEEEauthorrefmark{1},
Ahmad Bazzi\IEEEauthorrefmark{1}\IEEEauthorrefmark{2}, 
and Marwa Chafii\IEEEauthorrefmark{1}\IEEEauthorrefmark{2}}

\IEEEauthorblockA{\IEEEauthorrefmark{1}Engineering Division, New York University (NYU) Abu Dhabi, Abu Dhabi, UAE}
\IEEEauthorblockA{\IEEEauthorrefmark{2}NYU WIRELESS, NYU Tandon School of Engineering, New York, USA \\
%Email: \{oscarwang, roberto.bomfin, ahmad.bazzi, marwa.chafii\}@nyu.edu
}
}

\maketitle

\begin{abstract}
Multi-band sensing has emerged as a key enabler of integrated sensing and communication (ISAC), one of the six primary usage scenarios defined for IMT-2030 (6G). The introduction of frequency range 3 (FR3, 7-24 GHz), comprising non-contiguous sub-bands across a wide frequency span, further reinforces the importance of multi-band operation. In such scenarios, frequency-dependent clutter, collectively referred to as dense multipath components (DMC), must be carefully considered. Building on prior literature and our experimental observations, this paper analyzes the impact of DMC on multi-band fusion ISAC systems by investigating Cram\'er-Rao bound (CRB)-based fundamental limits and the performance of our proposed multi-band estimator. Numerical results show that multi-band processing, especially in DMC-dominated scenarios, can substantially reduce estimation error and boost system resilience when channel statistics vary.
\end{abstract}

\begin{IEEEkeywords}
multi-band sensing, integrated sensing and communication, localization, dense multipath components, ISAC, FR3, 6G
\end{IEEEkeywords}

\section{Introduction}
\label{sec:introduction}
Future 6G systems are envisioned to reliably facilitate novel applications that interact more deeply with the physical world, such as the internet of things, autonomous driving, extended reality, smart cities, digital twins, and radio maps for environment-aware communication. Thus, to save resources in bandwidth, hardware, processing power, and energy while enabling the efficient operation of such a large number of devices, \ac{ISAC} has been proposed as one of the six key usage scenarios of IMT-2030 (6G)\cite{chafii2023twelve}, which investigates sensing via existing communication protocols and infrastructure \cite{11161464}, including the allocated frequency bands. With the addition of frequency range 3 (FR3, 7-24 GHz) \cite{bazzi2025uppermidbandspectrum6g}, future 6G systems will likely operate in a wide range of frequencies and even multiple sub-bands \cite{Mezzavilla2024frequencyhopping}. Hence, performing multi-band sensing in FR3 using communication signals is naturally a promising enabler of \ac{ISAC} \cite{raviv2024multi,wang2025compressed}, given the potential of sensing capabilities for 6G \cite{yu2026sensing}.

Multi-band sensing is the technique of performing sensing across possibly non-contiguous frequency bands. According to the literature, some reasons for choosing multi-band sensing include the following. First, although using a large, contiguous bandwidth is a typical and intuitive approach to achieving higher time resolution, it is associated with increased hardware and algorithmic complexity \cite{hu2023mbrhs}. Moreover, communication-centric \ac{ISAC} systems are typically unable to process ideal radar-only wideband sensing waveforms, such as radar chirps \cite{wan2024ofdm}. Additionally, the increased amount of sensing data implies a lower error probability, such as when severe fading happens at a particular sub-band. Finally, some spectrum, such as FR3, is highly fragmented due to coexistence with incumbent applications \cite{wan2024ofdm}, such as satellites \cite{testolina2024sharing}. Multi-band sensing is thus a reasonable choice for using the wide bandwidth in FR3. 

Typical sensing processing in \ac{ISAC} literature relies on parameter estimation from \ac{SC}-only channel models defined by target geometry \cite{Bomfin_Chafii_UW_ISAC_2024}. However, an important propagation effect, well studied in channel modeling but largely overlooked in \ac{ISAC}, is the \ac{DMC} \cite{molish2022DMCSurvey, Richter_DMC_Tracking_2006}. \ac{DMC} refers to interference due to clutter such as diffuse scattering and diffraction, which can come from the background environment (walls, building structure, etc.). Neglecting \ac{DMC} leads to overly optimistic predictions of sensing performance or to non-negligible estimation bias and errors, yielding erroneous internal reliability metrics. 

This issue is exacerbated in multi-band settings, where \ac{DMC} exhibits frequency-dependent statistics, as observed in our experiment-based works \cite{bomfin2024experiment,bomfin2026icc}. This effect induces nonlinear behavior that invalidates conventional intuitions and necessitates \ac{DMC}-aware modeling to effectively employ and accurately assess multi-band \ac{ISAC} systems. Despite its importance, the role of \ac{DMC} in multi-band sensing remains poorly understood. For example, multi-band sensing works such as \cite{wan2024ofdm,hu2023mbrhs} all assumed an \ac{SC}-only channel model with white Gaussian noise, while \ac{DMC}-related works such as \cite{molish2022DMCSurvey,  Richter_DMC_Tracking_2006} are focused on \ac{CM} and single-band parameter estimation. Our previous works \cite{bomfin2024experiment,bomfin2026icc} focused on single-band processing, whereas joint multi-band processing was left for future work. The theoretical framework for multi-band \textit{fusion} processing with sub-band frequency-dependent \ac{DMC} remains an open problem. 

In this paper, we bridge the gap between multi-band sensing and DMC. First, we define the \ac{DMC}-affected system model in Section \ref{sec:system-model}. Next, we derive two important fundamental limits: the \ac{CRB} and the \ac{ESNR} in Section \ref{sec:fundamental-limits}. To operationalize our insights, we then explain in Section \ref{sec:proposed-multi-band-estimation-algorithm} how the fundamental limits can be achieved through our proposed multi-band estimation algorithm. Finally, we provide an in-depth analysis of our theoretical and algorithmic numerical results in Section \ref{sec:numerical-results}. Our contributions in this paper are summarized as follows:
\begin{itemize}
\item We derive and investigate the \ac{CRB}-based fundamental sensing limits, and show how our proposed multi-band estimation algorithm can achieve it.
\item We show that \ac{DMC}-induced nonlinearities challenge conventional sensing intuitions, in which one sub-band no longer dominates performance across all scenarios.
\item We show how multi-band processing can improve estimation accuracy, especially in \ac{DMC}-dominated scenarios.
\item We show how multi-band processing can enhance the resilience of system performance to variations in channel statistics in \ac{DMC}-dominated environments.
\end{itemize}

\textbf{Notations}: $\otimes$ denotes the Kronecker product, $\odot$ denotes the element-wise product, and $\toep(\pmb{x})$ denotes the Toeplitz matrix where the first column is $\pmb{x}$ and the first row is $\pmb{x}^H$. 
Moreover, $\cat_n\{\cdot\}$ denotes concatenating tensors in the set $\{ \cdot \}$ along the $n^{\text{th}}$ dimension.
Also, $\{ a_n \}_{n=1}^{N}$, which may also be written as $a_{1:N}$, denotes the set $\{ a_1,a_2,\dots,a_N \}$, and $[ \pmb{a}_n ]_{n = 1}^N$ denotes the vertically stacked vector $[\pmb{a}_1^T,\pmb{a}_2^T,\dots,\pmb{a}_N^T ]^T$.
Additionally, $\pmb{0}_{\times n}$ denotes a matrix of all zeros with $n$ columns of a suitable number of rows.
Finally, for some matrix $\pmb{A}$, $\left[\pmb{A}\right]_{i,:}$ denotes the $i^{\text{th}}$ row, $\left[\pmb{A}\right]_{:,j}$ denotes the $j^{\text{th}}$ column, and $\left[\pmb{A}\right]_{i,j}$ denotes the $(i,j)^{\text{th}}$ element.

\section{System Model}
\label{sec:system-model}
\begin{figure}[!t]
\centering
\includegraphics[width=3.5in]{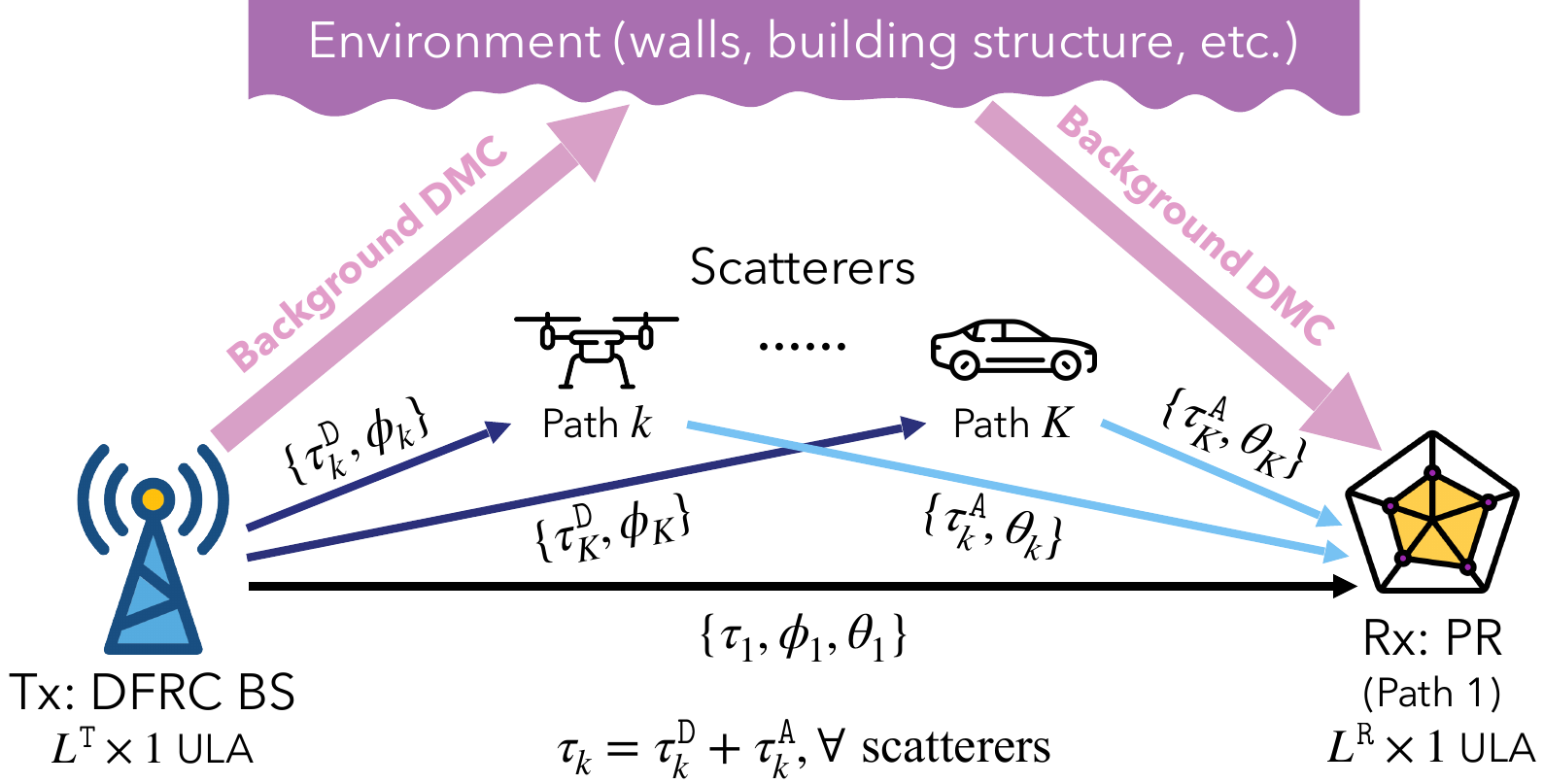}
\vspace{-10pt}
\caption{Bi-static ISAC scenario with $K$ paths under background \ac{DMC} interference. The geometric sensing parameters associated with each path are shown.}
\label{fig:scenario}
\vspace{-10pt}
\end{figure}

The scenario, as shown in Fig. \ref{fig:scenario}, involves a \ac{DFRC} \ac{BS} serving as the \ac{Tx} and a \ac{PR} as the \ac{Rx}, whose Cartesian locations are assumed to be known. The \ac{Tx} \ac{DFRC} \ac{BS} has $L^{\tt{T}}$ \ac{ULA} antenna elements, and the \ac{Rx} \ac{PR} has $L^{\tt{R}}$ \ac{ULA} antenna elements.  The \ac{PR} receives the multi-band signal from the \ac{BS} across $M$ sub-bands (indexed by $m$), each with $N$ subcarriers (indexed by $n$), and performs bistatic sensing using downlink preamble signals from communications with user equipment in the environment. In this paper, all sub-bands are assumed to have the same number of subcarriers, even though this can be easily generalized. The channel consists of $K$ paths (indexed by $k$, where $k=1$ is the \ac{LoS} path) and the \ac{DMC}. 
We assume each path is specular, meaning the only source of \ac{DMC} is the background environment.

In this paper, we study the typical \ac{ISAC} system in which sensing is performed on the \ac{MIMO} frequency-domain \ac{CSI}. In practice, the \ac{CSI} is obtained via \ac{OFDM} pilots that are designed to ensure orthogonality across \ac{Tx} antennas. Following the model in \cite{molish2022DMCSurvey}, we express the multi-band \ac{MIMO} channel \ac{TF} $\pmb{h} \in \mathbb{C}^{NL^{\tt{T}}L^{\tt{R}}M \times 1}$ by $\pmb{h} = \pmb{s} + \pmb{d} + \pmb{w}$,
where $\pmb{s} = \left[{\pmb{s}_{m}}\right]_{m=1}^M$ represents the \ac{SC} associated with the $K$ paths, $\pmb{d}\sim\mathcal{CN}\left(\pmb{0},\pmb{R}\right)$ represents the background \ac{DMC}, and $\pmb{w}\sim\mathcal{CN}\left(\pmb{0},\pmb{S}\right)$ represents the noise with covariance matrix $\pmb{S}$. The detailed modeling for each of the three parts (\ac{SC}, \ac{DMC}, and noise) is discussed in the following subsections.

Our problem definition for this work is: given the background \ac{DMC}-affected multi-band \ac{MIMO} channel \ac{TF} $\pmb{h}$ at the \ac{Rx}, we jointly estimate the sensing parameters, including the number of paths $K$, the geometric parameters of the paths, namely \acp{AoA}, \acp{AoD}, \acp{ToA}, and path gains. 

\subsection{The Specular Components (SC)}

The \ac{SC} is considered to be deterministic. Hence, each $\pmb{s}_{m} \in \mathbb{C}^{NL^{\tt{T}}L^{\tt{R}} \times 1}$ represents the specular channel \ac{TF} for each sub-band with frequency $f_m$, which is given by
\begin{equation}
\label{eq:h_per_subcarrier}
\pmb{s}_{m} = 
\sum_{k=1}^K g_{k,m} \pmb{a}_{m}(\tau_k,\phi_k,\theta_k),
\end{equation}
where $\pmb{a}_{m}(\tau_k,\phi_k,\theta_k)$ is the total steering vector for path $k$ at sub-band $m$ described by the geometric parameters, and $g_{k,m}$ is the path gain for path $k$ at sub-band $m$.

The geometric parameters of the $k^{\textrm{th}}$ path include the \ac{ToA} delay $\tau_k$, the azimuth \ac{AoD} $\phi_k$ relative to the \ac{Tx} array broadside, and the azimuth \ac{AoA} $\theta_k$ relative to the \ac{Rx} array broadside. They contribute to the \ac{SC} part of the model through the total steering vector, which can be factorized as \cite{molish2022DMCSurvey}
\begin{equation}
\pmb{a}_{m}(\tau_k,\phi_k,\theta_k) = 
\pmb{a}^{\tt{F}}_m(\tau_k) \otimes 
\pmb{a}^{\tt{T}}_m(\phi_k) \otimes 
\pmb{a}^{\tt{R}}_m(\theta_k).
\end{equation}
Assuming typical far-field propagation and narrowband antenna array response \cite{molish2022DMCSurvey}, the factorized terms of the \ac{SC} steering vector can be expressed as
\begin{equation}
\label{aF}
	\pmb{a}^{\tt{F}}_m(\tau) = 
	\left[
		\exp\left(
			-jn\omega_{\Delta,m} \tau
		\right)
	\right]_{n=0}^{N-1},
\end{equation}
\begin{equation}
\label{aT}
	\pmb{a}^{\tt{T}}_m(\phi) =
	\left[
		\exp\left(
			{-j \omega_m (d^{\tt{T}}/c)}
			\ell
			\sin\phi
		\right)
	\right]_{\ell = 0}^{L^{\tt{T}}-1},
\end{equation}
\begin{equation}
\label{aR}
	\pmb{a}^{\tt{R}}_m(\theta) =
	\left[
		\exp\left(
			{-j \omega_m (d^{\tt{R}}/c)}
			\ell
			\sin\theta
		\right)
	\right]_{\ell = 0}^{L^{\tt{R}}-1},
\end{equation}
where $\omega_{m} = 2\pi f_m$, $\omega_{\Delta,m} = 2\pi f_{\Delta,m}$ is the subcarrier spacing, and $d^{\tt{T}}$ and $d^{\tt{R}}$ are the antenna spacings of the \ac{Tx} and \ac{Rx} \ac{ULA}, respectively. 

The path gains, which are frequency-dependent due to antenna gains and complex path coefficients, can be modeled using the Friis and bistatic radar equations. The path gains are also dependent on $\tau^{\tt{D}}_k$, the delay between the \ac{Tx} and the scatterer, $\tau^{\tt{A}}_k$, the delay between the scatterer and the \ac{Rx}, and $P^{\tt{T}}$, the fixed transmit power spectral density across symbols. The average transmit power per subcarrier across symbols on the $m^{\textrm{th}}$ sub-band is thus $P^{\tt{T}} f_{\Delta,m}$.

\subsection{The Dense Multipath Components (DMC)}

Since \ac{DMC} arises from small surface details such as roughness, it may exhibit randomness across channel realizations due to small-scale mobility on the order of wavelengths. We employ the widely accepted \ac{DMC} model in \cite{molish2022DMCSurvey,rimaxthesis}. \ac{DMC} profiles supporting this modeling approach are observed and measured in our experiment-based works \cite{bomfin2024experiment,bomfin2026icc}.

\subsubsection{Delay domain characteristics}
The \ac{PDP} of the complex Gaussian process can be modeled as an exponential decay \cite{rimaxthesis}:
\begin{equation}
P_m^{\tt{DS}}(\tau,\tau_1) =
\begin{cases} 
0 & \tau < \tau_1, \\
\alpha \magn{ g_{1,m} }^2 / 2 & \tau = \tau_1, \\
\alpha \magn{ g_{1,m} }^2 e^{-\beta_m(\tau - \tau_1)} & \tau > \tau_1,
\end{cases}
\end{equation}
where the maximum power of the \ac{DMC} is modeled as a fraction of the \ac{LoS} \ac{SC} power quantified by $\alpha$, and $\beta_m$ is the decay rate for the $m^{\textrm{th}}$ sub-band.

In the frequency domain, $\beta_m$ can also be interpreted as the coherence bandwidth, and the \ac{DMC} sampled at the subcarriers can be modeled as joint complex Gaussian random variables following the distribution $\mathcal{CN}(\pmb{0},\pmb{R}_m^{\tt{F}})$, where $\pmb{R}_m^{\tt{F}} \in \mathbb{C}^{N\times N}$ is given by $\pmb{R}_m^{\tt{F}} = \toep(\pmb{r}^{\tt{F}}_m)$, where
\begin{equation}
\pmb{r}^{\tt{F}}_m = 
\frac{\alpha \magn{ g_{1,m} }^2}{N} 
\left[
 \frac{e^{-j n \omega_{\Delta,m} \tau_1}}{\tilde{\beta}_m + \frac{j 2\pi n}{N}}
\right]_{n=0}^{N-1},
\end{equation}
and $\tilde{\beta}_m = \beta_m/(N-1)\Delta\omega_m$ is the coherence bandwidth for the $m^{\textrm{th}}$ sub-band, normalized to the measurement bandwidth $(N-1)\Delta\omega_m$.

\subsubsection{Angle domain characteristics}

In this paper, the angular covariance matrices of the \ac{DMC} $\pmb{R}^{\tt{T}}_m$ and $\pmb{R}^{\tt{R}}_m$ are modeled as $\pmb{I}$. This is because exact modeling of these matrices would require knowledge of the system-specific frequency-dependent antenna patterns, which is more suitable in works involving experimentation \cite{bomfin2024experiment,bomfin2026icc}, and this is planned for future work. Thus, we use the typical assumption of white \ac{DMC} in space, without any specific \ac{PAP} and antenna patterns, which is more tractable and sufficient for deriving the most important insights of the paper.

Now, based on the above definitions and assuming independence across different sub-bands, we can compactly write the \ac{DMC} covariance matrix as
$\pmb{R} = \blockdiag\left\{\pmb{R}_m\right\}_{m=1}^M$,
where
\begin{equation}
\pmb{R}_m = 
\pmb{R}_m^{\tt{F}}
\otimes\pmb{R}_m^{\tt{T}}
\otimes\pmb{R}_m^{\tt{R}}
	\in
	\mathbb{C}^{NL^{\tt{T}}L^{\tt{R}} \times NL^{\tt{T}}L^{\tt{R}}}.
\end{equation}

\subsection{The Noise}
We assume the noise is white; thus, the noise covariance matrix is modeled as
\begin{equation}
	\pmb{S} = \blockdiag{\left\{\pmb{S}_m\right\}}_{m=1}^M
	= \blockdiag{\left\{\sigma_m^2\pmb{I}\right\}}_{m=1}^M,
\end{equation}
and $\sigma_m^2 = \NF\cdot N_0 f_{\Delta,m}$ is the per-subcarrier noise variance for sub-band $m$, where $N_0$ is the white Gaussian noise power spectral density and $\NF$ is the noise figure. 

In this work, our goal is to investigate the impact of \ac{DMC} on multi-band sensing. Therefore, communication–sensing trade-offs in channel estimation are not incorporated into the model. We remark that when channel estimation is performed using payload data instead of pilots, the resulting estimation error is generally colored rather than white. Nevertheless, the conclusions of this study remain valid, since our model accommodates an arbitrary noise covariance matrix.

\section{Fundamental Limits}
\label{sec:fundamental-limits}
The \ac{CRB}-based fundamental limits discussed in this section are crucial for providing intuitive and novel insights into multi-band sensing by capturing the interplay among \ac{SC}, \ac{DMC}, and noise effects simultaneously. They are also helpful for designing estimation algorithms by signaling the performance bounds. In this paper, the \ac{CRB} assumes perfect knowledge of the noise figure and the \ac{DMC} parameters $\alpha$ and $\beta_{1:M}$.

\subsection{Cram\'er-Rao Bound (CRB)}
Assuming independence between the background \ac{DMC} and Gaussian noise, the distribution of $\pmb{h}$ can be modeled as $\pmb{h} \sim \mathcal{CN}\left(\pmb{s},\pmb{M} \right)$, where
\begin{equation}
\pmb{M} = \blockdiag\left\{\pmb{M}_m\right\}_{m=1}^M = \blockdiag\left\{\pmb{R}_m+\pmb{S}_m\right\}_{m=1}^M
\end{equation}
is known in the context of this paper.
Hence, exploiting the block-diagonal structure of $\pmb{M}$, following the general complex Gaussian \ac{FIM} structure in \cite{kay1993estimation}, the \ac{FIM} is constructed as follows
\begin{equation}
\label{FIM}
	\pmb{F} = \sum_{m=1}^M \pmb{F}_m = \sum_{m=1}^M 2 \Re\left({\pmb{D}_m^H\pmb{M}_m^{-1}\pmb{D}_m}\right),
\end{equation}
where $\pmb{D}_m$ is the Jacobian formed by the partial derivatives of the mean vector $\pmb{s}$ (the \ac{SC}) with respect to the sensing parameters on each sub-band. 
The \ac{CRB} of the $i^{\text{th}}$ parameter can hence be obtained by $\CRB_{i} = \left[\pmb{F}^{-1}\right]_{i,i}$. 

We now derive expressions for the Jacobian matrix $\pmb{D}$ of the \ac{FIM}. 
We need
\begin{equation}
	\pmb{D}_m = 
	\cat_2\left\{
		\pmb{D}_m^{\tau},
		\pmb{D}_m^{\phi},
		\pmb{D}_m^{\theta},
		\pmb{D}_m^{\Re g},
		\pmb{D}_m^{\Im g}
	\right\}.
\end{equation}
First, for $\pmb{s}$ with respect to the geometric parameters, we have
\begin{equation}
\label{DsToA}
	\pmb{D}_m^{\tau} = 
	\cat_2\left\{
	g_{k,m} \frac{\partial}{\partial \tau_k}
	\pmb{a}_m (\tau_k,\phi_k,\theta_k)
	\right\}_{k=1}^K,
\end{equation}
and $\pmb{D}_{\phi}$ and $\pmb{D}_{\theta}$ follow similar structures from \eqref{DsToA} by changing $\tau$ to $\phi$ and $\theta$. Due to the Kronecker structure, we only need to know the following:
\begin{equation}
\label{partialaFToA}
	\frac{\partial \pmb{a}^{\tt{F}}_m}{\partial \tau_k} = 
	\pmb{a}^{\tt{F}}_m \odot 
	\left[
		-jn\omega_{\Delta,m}
	\right]_{n=0}^{N-1},
\end{equation}
\begin{equation}
\label{partialaTxAoD}
	\frac{\partial \pmb{a}^{\tt{T}}_m}{\partial \phi_k} = 
	\pmb{a}^{\tt{T}}_m \odot
	\left[
		-j\omega_m  (d^{\tt{T}}/c) \ell \cos \phi_k
	\right]_{\ell=0}^{L^{\tt{T}}-1},
\end{equation}
\begin{equation}
\label{partialaRxAoA}
	\frac{\partial \pmb{a}^{\tt{R}}_m}{\partial \theta_k} = 
	\pmb{a}^{\tt{R}}_m \odot
	\left[
		-j\omega_m  (d^{\tt{R}}/c) \ell \cos \theta_k
	\right]_{\ell=0}^{L^{\tt{R}}-1},
\end{equation}
Then, for $\pmb{s}$ with respect to the gains, we have
\begin{equation}
	\pmb{D}_m^{\Re g}
	=
	\cat_2\left\{
	\left[
	\pmb{0}_{\times (m-1)}, 
		\pmb{a}_m(\tau_k,\phi_k,\theta_k), 
	\pmb{0}_{\times (M-m)}
	\right]
	\right\}_{k=1}^K,
\end{equation}
\begin{equation}
	\pmb{D}_m^{\Im g}
	=
	\cat_2\left\{
	\left[
	\pmb{0}_{\times (m-1)}, 
		j\pmb{a}_m(\tau_k,\phi_k,\theta_k), 
	\pmb{0}_{\times (M-m)}
	\right]
	\right\}_{k=1}^K.
\end{equation}
Note that although the gains $g_{k,m}$ are modeled with dependence on the bistatic delays $\tau_k^{\tt{D}}$ and $\tau_k^{\tt{A}}$, our algorithm estimates the gains and delays independently. Thus, our \ac{CRB} formulation also ignores the dependence of the gains on the delays.
 
\subsection{Estimation Signal-to-Noise Ratio (ESNR)}

The \ac{ESNR} can be thought of as a measure of the certainty of an estimated path. It can be used to determine whether a peak is \ac{SC} or part of the \ac{DMC} profile, and whether an estimated path can be trusted.
For a path with gain $g_{k,m}$, the \ac{ESNR} $\mu_{k,m}$ is defined as
\begin{equation}
	\mu_{k,m} = \frac{\lvert g_{k,m} \rvert^2}{\CRB_{\lvert g_{k,m} \rvert}},
\end{equation}
where $\CRB_{\lvert g_{k,m} \rvert}$ can be obtained from $\CRB_{\Re g_{k,m}}$ and $\CRB_{\Im g_{k,m}}$ through a Jacobian coordinate transformation on the \ac{FIM}. The \ac{ESNR} plays a crucial role in implementing the multi-band estimation algorithm described in Section \ref{sec:proposed-multi-band-estimation-algorithm}.

\section{Proposed Multi-Band Estimation Algorithm}
\label{sec:proposed-multi-band-estimation-algorithm}
\begin{figure}[!t]
\centering
\includegraphics[width=3.4in]{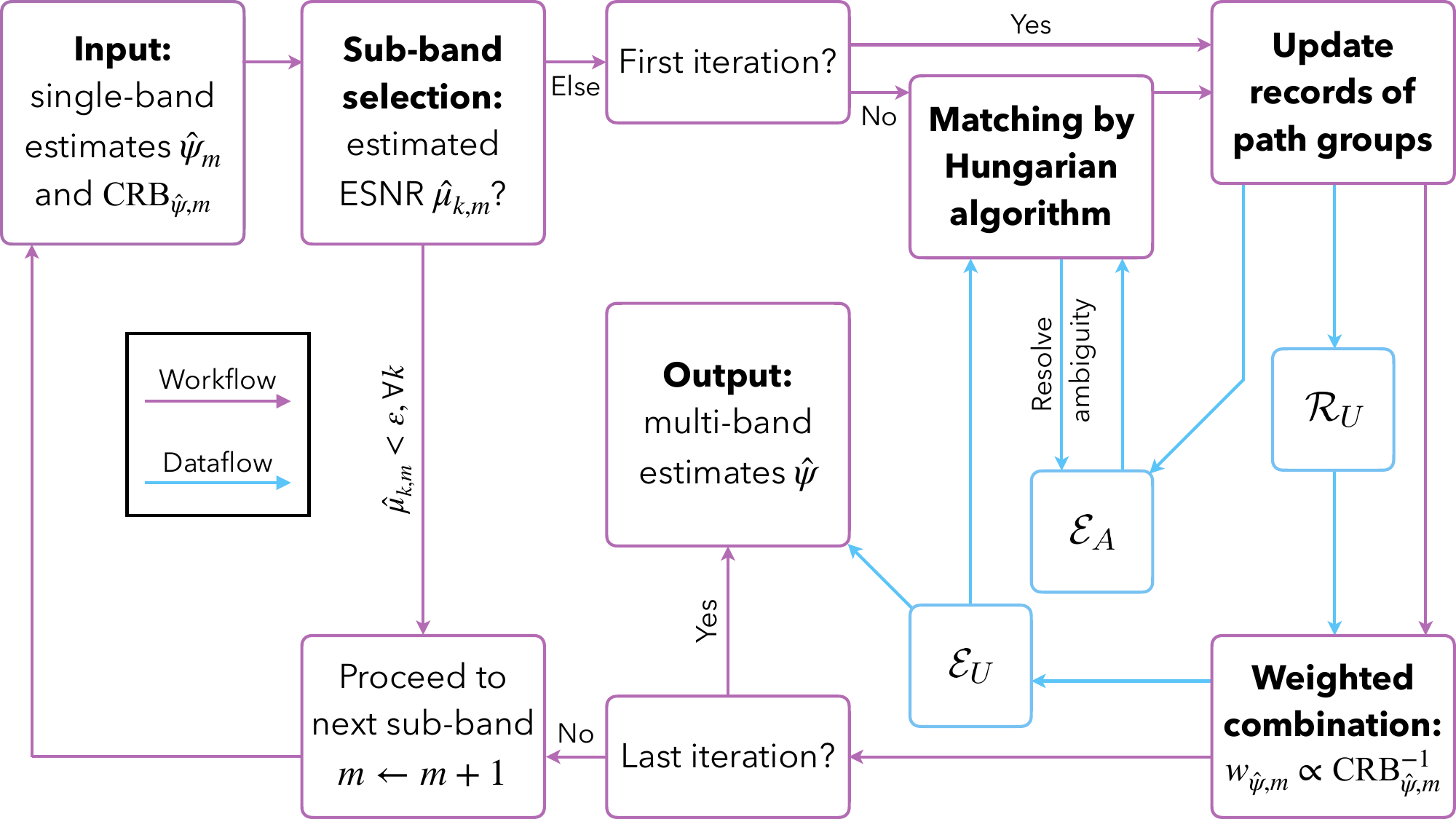}
\caption{Flowchart of the key components and steps of the algorithm.}
\label{fig:flowchart}
\vspace{-10pt}
\end{figure}

In this section, we explain the essence of our proposed scalable multi-band estimator based on fusion of single-band estimators in \cite{wang2026mbdmc}. Our goal is to derive a tractable near-optimal estimator. Instead of solving the full multi-band estimation problem jointly, we combine results per sub-band with minimal performance loss, which considerably reduces complexity. A flowchart summary of the algorithm is provided in Fig. \ref{fig:flowchart}, where $\hat{\psi}$ refers to the estimate for any geometric parameter $\psi$ (including \ac{ToA} $\tau$, \ac{AoD} $\phi$, or \ac{AoA} $\theta$).
For the single-band estimators, we choose the well-known state-of-the-art estimation algorithm described in \cite{rimaxthesis} based on the iterative \ac{LM} approach, which is known to achieve the \ac{CRB} in high \ac{SNR}. 

\subsection{Multi-Band Fusion}

To easily accommodate multi-band operation, we wish to perform a linear combination of the estimated geometric parameter values (\acp{ToA}, \acp{AoA}, and \acp{AoD}) from the single-band estimators. When the estimators for every sub-band are efficient (i.e., they approach their own \ac{CRB}), we can obtain the minimal variance linear unbiased estimator by designing weights for any sub-band $m$ as
\begin{equation}
\label{weights}
	w_{\hat{\psi}_k,m} = \frac{\CRB^{-1}_{\hat{\psi}_k,m}}
	{\sum_{\forall m \text{ detected}} \CRB^{-1}_{\hat{\psi}_k,m}}
\end{equation}
To determine how certain we can be about a path seen by a given sub-band, we use the estimated \ac{ESNR}, computed from the estimated path gains and their respective \acp{CRB}, to perform sub-band selection.
 If $\hat{\mu}_{k,m} < \varepsilon_{\tt{ESNR}}, \forall k$, where $\varepsilon_{\tt{ESNR}}$ is the predefined \ac{ESNR} threshold, the sub-band is uncertain and the path should be dropped for combining. 

\subsection{Resolving Angular Ambiguity}

Since the antenna spacings $d^{\tt{T}}$ and $d^{\tt{R}}$ are fixed across frequencies, grating lobes (spatial aliasing) may occur when $d > \lambda_m/2$, yielding multiple angle estimates for the same physical path. We consider all these aliased versions as a \textit{path group}.
To resolve this ambiguity, we apply the Hungarian matching algorithm across sub-bands. The matching cost between one of the latest estimated path groups $p$ and one of the newly estimated ones $i$ at sub-band $m$ is 
\begin{equation}
\left[{\pmb{C}}_{m}\right]_{p,i} = \min\limits_{r_{1:4}}
    \norm{
    \mathcal{T}\left( \hat{\pmb{x}}_p^{r_1,r_2}\right) - 
    \mathcal{T}\left(\hat{\pmb{x}}_i^{r_3,r_4}\right)}_2,
    \end{equation}
where $r_{1:4}$ indexes all possible ambiguity versions, $\hat{\pmb{x}}_p^{r_1,r_2} = [\hat{\tau}_p, \hat{\phi}^{r_1}_p, \hat{\theta}^{r_2}_p]$, $\hat{\pmb{x}}_i^{r_3,r_4} = [\hat{\tau}_i, \hat{\phi}^{r_3}_i, \hat{\theta}^{r_4}_p]$, and $\mathcal{T}(\cdot)$ is a normalization mapping explained in detail in \cite{wang2026mbdmc}.
    A match is accepted if its cost is below $C_{\tt{max}}$ and if it is sufficiently prominent, i.e., the selected cost is at least $\varepsilon_{\tt{prom}}$ smaller than the second-best candidate across all possible ambiguity versions.

The proposed multi-band estimator iterates over every sub-band $m$, and it dynamically keeps and updates, per valid iteration, three data containers of all estimated path groups: 
\begin{itemize}
	\item $\mathcal{E}_U$: the latest weighted combined estimates of the unambiguous paths.
	\item $\mathcal{E}_A$: the latest estimates of the ambiguous path groups.
	\item $\mathcal{R}_U$: the record containing all the past estimates of the unambiguous paths.
\end{itemize}

\section{Numerical Results}
\label{sec:numerical-results}
\begin{table}[!t]
\renewcommand{\arraystretch}{1.3}
\caption{Numerical Parameters}
\label{table:numerical-parameters}
\centering
\newcolumntype{Y}{>{\centering\arraybackslash}m{79.125pt}}
\begin{tabularx}{\columnwidth}{|c|Y||c|Y|}
\hline
\rowcolor{gray10pct} \multicolumn{4}{|c|}{\bfseries System Parameters} \\
\hline
$M$ & 2 & $N$ & 128 \\
\hline
$L^{\tt{T}}$ & 2 & $L^{\tt{R}}$ & 2 \\
\hline
$d^{\tt{T}}$ & 0.02 m & $d^{\tt{R}}$ & 0.02 m \\
\hline
$f_m$ & $\{8.75,21.7\} \text{ GHz}$ & $f_{\Delta,m}$ & $\{1000,1000\} \text{ kHz}$ \\
\hline
$\BW_m$ & $\{128,128\} \text{ MHz}$ & $K$ & 2  \\
\hline
$N_0$ & -174 dBm/Hz & $\NF$ & 7 dB \\
\hline
\rowcolor{gray10pct} \multicolumn{4}{|c|}{\bfseries Algorithm Parameters} \\
\hline
$\varepsilon_{\tt{ESNR}}$ & 6 dB & $\varepsilon_{\tt{prom}}$ & 0.2 \\
\hline
$N_{\tt{MC}}$ & 1024 & $C_{\tt{max}}$ & 0.75 \\
\hline
\end{tabularx}
\vspace{-10pt}
\end{table}

We aim to illustrate not only the expected tradeoff among the \ac{SC}, \ac{DMC}, and noise, but also the gains of multi-band processing in terms of estimation accuracy, detection probability, and system resilience. A table of numerical parameters used is given in Table \ref{table:numerical-parameters}, where $N_{\tt{MC}}$ is the number of Monte Carlo simulations and $\BW_m$ is the bandwidth of the $m^{\textrm{th}}$ sub-band. The number of sub-bands, antennas, and paths are kept at 2 to clearly isolate the \ac{DMC}-induced trends.

\subsection{System Model Analysis}

\begin{figure}[!t]
\centering
\includegraphics[width=3.0in]{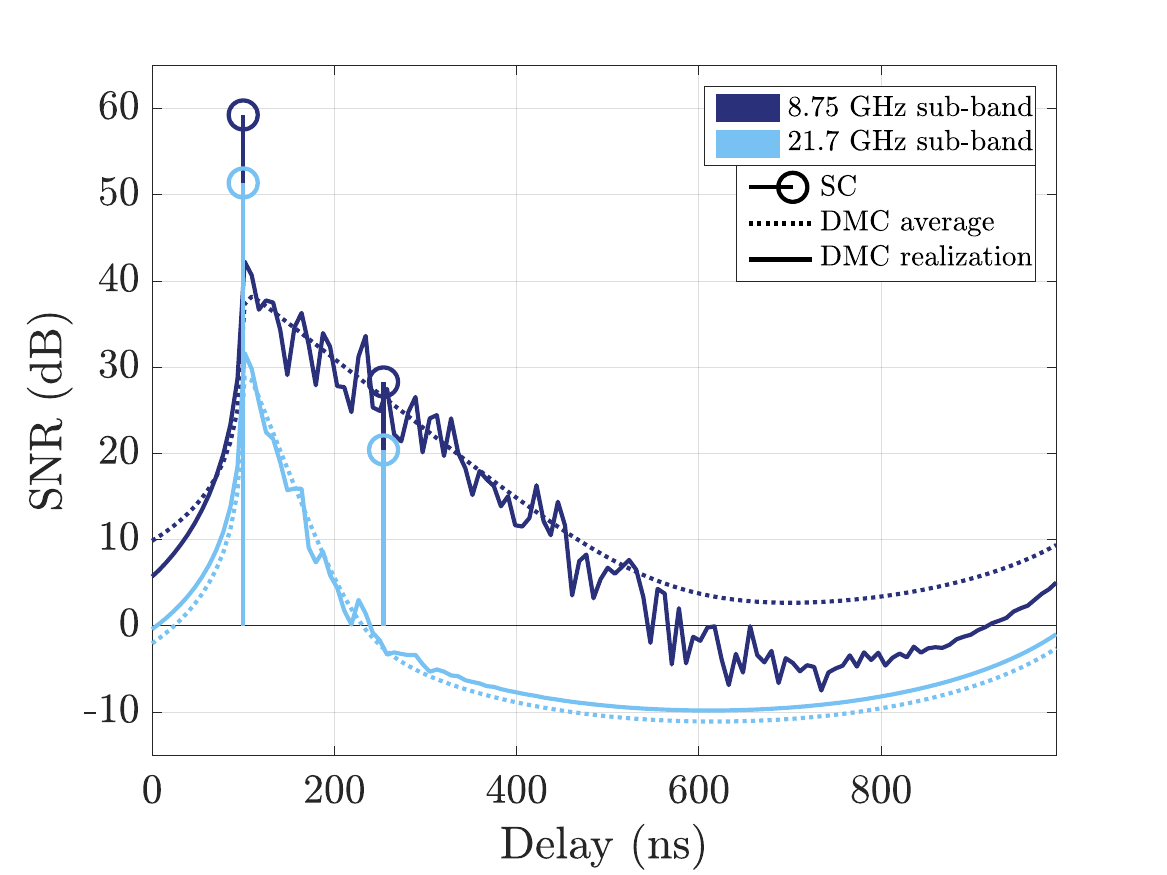}
\vspace{-5pt}
\caption{The \ac{PDP} of a two-path channel, including both \ac{DMC} and \ac{SC}, for both sub-bands at $P^{\tt{T}} = -40$ dBm/Hz, $\alpha = -20$ dB, and $\tilde{\beta}_m = \{0.15,0.5\}$ for the 8.75 GHz and 21.7 GHz sub-bands, respectively.}
\label{fig:pdp}
\vspace{-10pt}
\end{figure}

Fig. \ref{fig:pdp} shows the \ac{PDP} of an example two-path channel, providing a visual illustration of how the \ac{DMC} is impairing the \ac{SC} estimation. Since \ac{DMC} is multipath interference, it suffers from higher diffuse scattering, diffraction, and blockage as sub-band frequency increases, just like the \ac{SC}. Hence, it is practical to assume higher decay rates $\tilde{\beta}$ for higher frequency sub-bands, which is also experimentally verified by our other works \cite{bomfin2024experiment,bomfin2026icc}. We observe that the \ac{SC} in certain sub-bands, such as the scatterer path for the 8.75 GHz sub-band in the figure, may be shadowed under the \ac{DMC} profile. Also, due to the randomness of the \ac{DMC} profile, it may be hard to distinguish \ac{SC} from \ac{DMC} in finite bandwidth \acp{PDP}. Finally, a key tradeoff for multi-band systems in \ac{DMC}-affected channels is shown: although higher sub-bands typically suffer from lower \ac{SC} powers due to the propagation effects and more directive antenna patterns \cite{wang2026mbdmc}, they are also blessed by the faster \ac{DMC} decay rates, which reduces interference.

\subsection{Multi-Band Estimator Performance}

\begin{figure}[!t]
\centering
\includegraphics[width=3.0in]{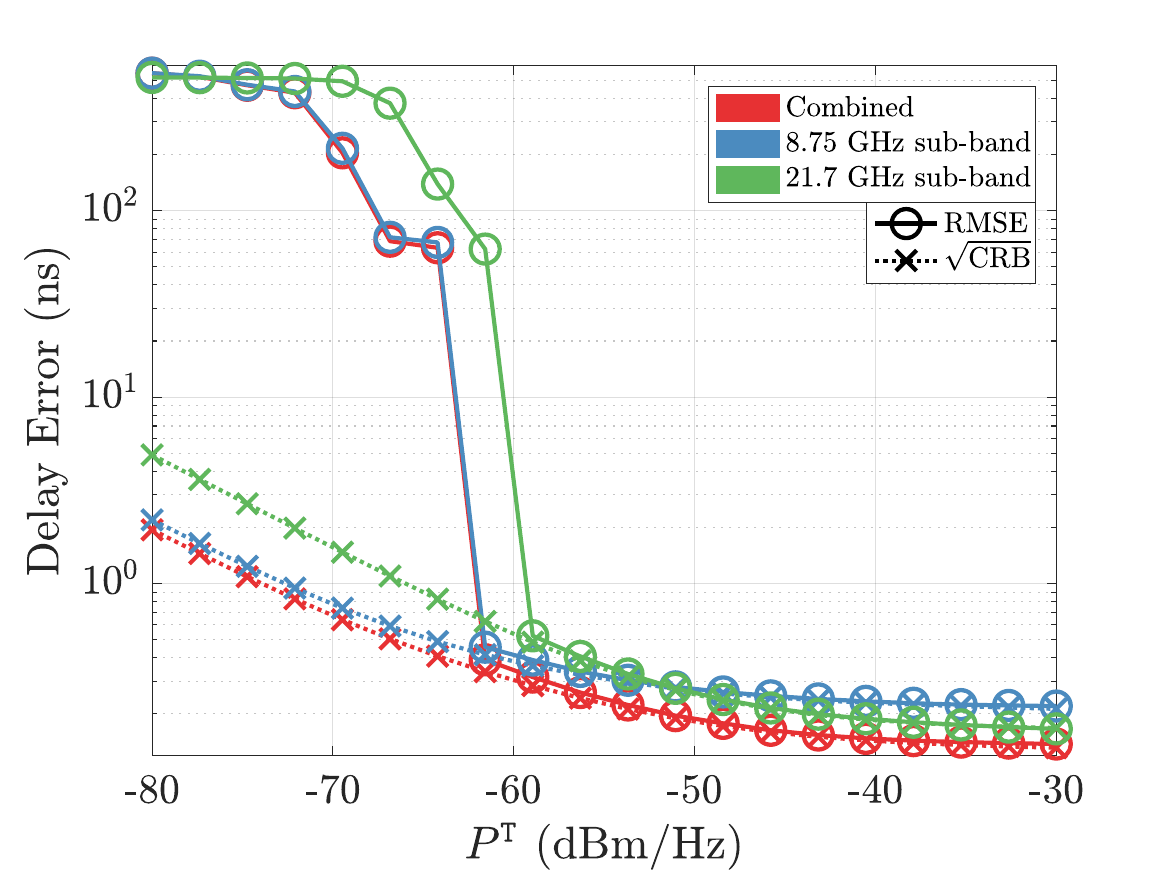}
\vspace{-5pt}
\caption{Numerical delay \ac{RMSE} and $\sqrt{\CRB}$ vs. $P^{\tt{T}}$ for a fixed scatterer in a two-path scenario, including both single- and multi-band cases, at $\alpha = -30$ dB and $\tilde{\beta}_m = \{0.5,1.5\}$.}
\label{fig:simulation_delay_error}
\vspace{-10pt}
\end{figure}

Fig. \ref{fig:simulation_delay_error} shows a simulation of the performance of the proposed estimation algorithm in a typical two-path scenario by comparing the estimation error in the \ac{ToA} of the scatterer against its \ac{CRB}. The \ac{Tx} is at $(8.99 \textrm{ m}, 0 \textrm{ m})$, the \ac{Rx} is at $(0 \textrm{ m}, 0 \textrm{ m})$, and the scatterer is fixed at $(3 \textrm{ m}, 1.8 \textrm{ m})$. First, we observe that the \ac{CRB} curves have a noise-dominated region (where the $\sqrt{\CRB}$ decreases linearly on a logarithmic scale) and a \ac{DMC}-dominated region (where the $\sqrt{\CRB}$ deviates from linear decrease and gradually approaches a floor). The reason behind this phenomenon is that the \ac{DMC} magnitude grows with $P^{\tt{T}}$. Moreover, we observe that across the two \ac{SNR} regions, the two sub-bands' performances ``switch" as a result of the previously mentioned key tradeoff in Fig. \ref{fig:pdp}. While ``switching" is not guaranteed in all cases, such a phenomenon being typical implies that one sub-band no longer dominates the estimation accuracy under all \ac{SNR} conditions, encouraging the use of multi-band sensing. 

Next, our proposed multi-band estimator achieves the \ac{CRB} beyond $P^{\tt{T}} = -61.58$ dBm/Hz by using both sub-bands, and the \ac{ESNR}-based sub-band selection scheme naturally switches the estimator to using the better performing 8.75 GHz sub-band below this \ac{SNR} level. Furthermore, the gain in \ac{RMSE} performance of the combined multi-band estimator, as compared to the performances of single-band estimators, is more significant in the \ac{CRB}-achieving region, especially in the ``switching" region. For example, at $P^{\tt{T}} = -30$ dBm/Hz, the delay \ac{RMSE} of the combined multi-band estimator is 0.14 ns, while those for the 8.75 GHz and 21.7 GHz single-band estimators are 0.22 ns and 0.17 ns, respectively. 

\subsection{Multi-Band Sensing Gains}

\begin{figure}[!t]
\centering
\includegraphics[width=3.0in]{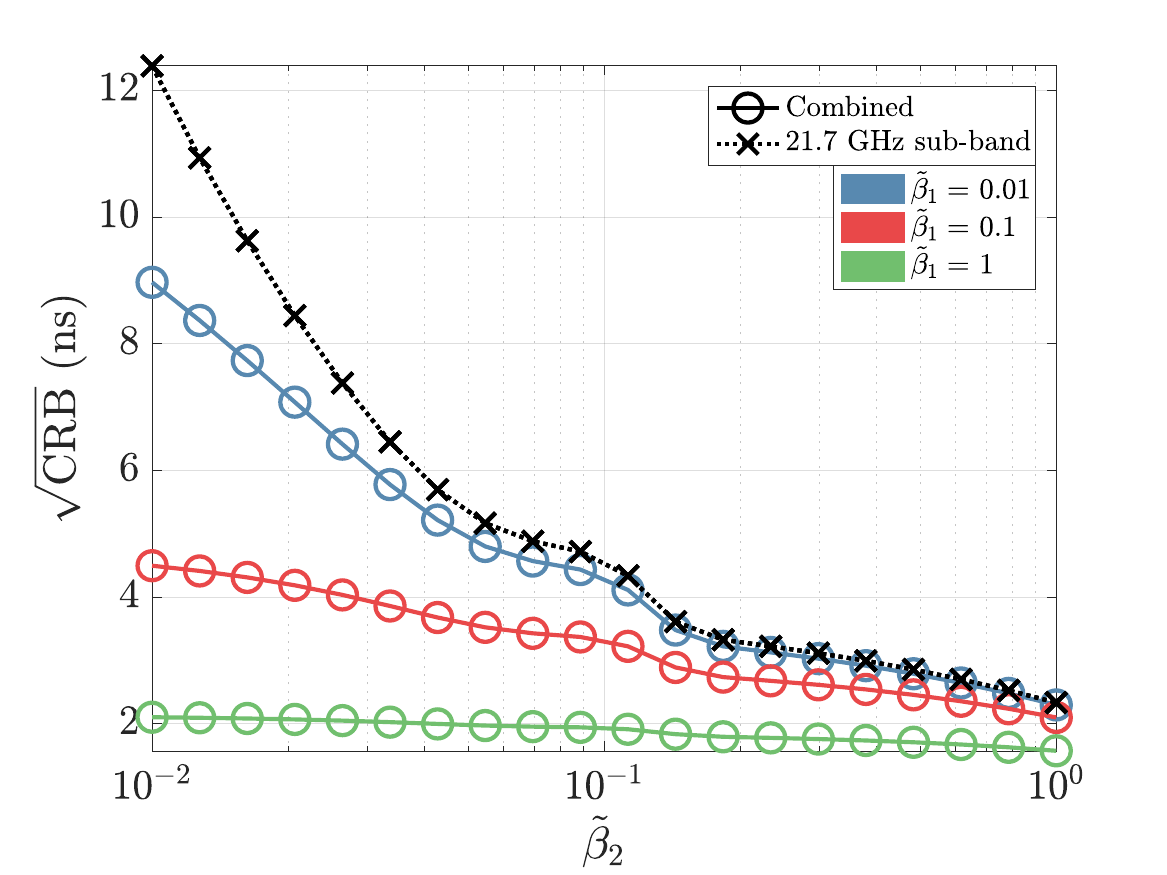}
\vspace{-5pt}
\caption{The delay $\sqrt{\CRB}$ vs. $\tilde{\beta}_2$ for a fixed scatterer in a two-path scenario, including both single- and multi-band cases, at $P^{\tt{T}} = -10$ dBm/Hz and $\alpha = -3$ dB, with multiple $\tilde{\beta}_1$ values.}
\label{fig:crb-beta2-var_beta1}
\vspace{-10pt}
\end{figure}

We are also interested in how multi-band sensing gains manifest as channel parameters, such as \ac{DMC} decay rates and scatterer locations, vary.
Fig. \ref{fig:crb-beta2-var_beta1} shows the \ac{CRB} of the \ac{ToA} of the scatterer in a scenario with the same \ac{Tx}, \ac{Rx}, and scatterer locations as in Fig. \ref{fig:simulation_delay_error}. We observe that the \ac{CRB} becomes more resilient to variations in the decay rate of this sub-band when being combined with another sub-band in a multi-band system, regardless of the decay rate of the other sub-band. This resilience, as well as the \ac{CRB} itself, improves with the decay rate of the other sub-band. The multi-band gain is more significant in regions with longer \ac{DMC} profiles. For example, when a 21.7 GHz sub-band is combined with an 8.75 GHz sub-band with $\tilde{\beta}_1 = 0.01$, the improvement in $\sqrt{\CRB}$ is only 0.04 ns if $\tilde{\beta}_2 = 1$, but increases to 3.42 ns if $\tilde{\beta}_2 = 0.01$, which translates to a significant error of 1.026 m.

\begin{figure}[!t]
\centering
\includegraphics[width=3.0in]{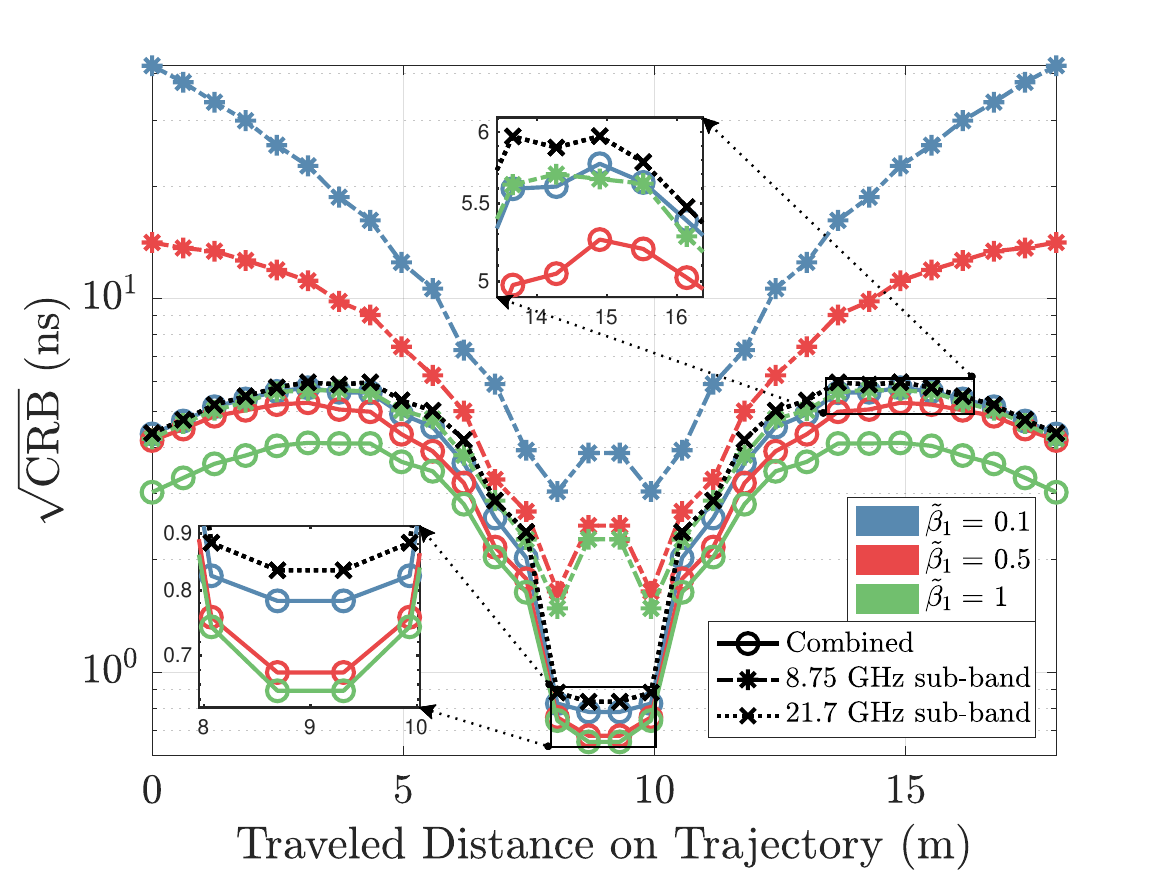}
\vspace{-5pt}
\caption{The delay $\sqrt{\CRB}$ of a scatterer along a trajectory in a two-path scenario, including both single- and multi-band cases, at $P^{\tt{T}} = -10$ dBm/Hz, $\alpha = -3$ dB, and $\tilde{\beta}_2 = 1$, with multiple $\tilde{\beta}_1$ values.}
\label{fig:crb-traj-var_beta1}
\vspace{-10pt}
\end{figure}

Fig. \ref{fig:crb-traj-var_beta1} shows the \ac{CRB} of the \ac{ToA} of the scatterer in a two-path scenario with the same \ac{Tx} and \ac{Rx} locations as in Fig. \ref{fig:simulation_delay_error}, while the scatterer moves along a trajectory from $(4.50 \textrm{ m}, 9 \textrm{ m})$ to $(4.50 \textrm{ m}, -9 \textrm{ m})$. When using only the 21.7 GHz sub-band, the variation of the $\sqrt{\CRB}$ across the entire trajectory is 5.13 ns; nevertheless, when combined with an 8.75 GHz sub-band with $\tilde{\beta}_1 = 0.5$, this variation reduces to 4.59 ns, and further down to 3.46 ns as $\tilde{\beta}_1$ increases to 1. Moreover, we observe that combining the 21.7 GHz sub-band with an 8.75 GHz sub-band with faster decay rates consistently improves the \ac{CRB} for all locations along the trajectory. We hence conclude that the benefits of multi-band fusion sensing in terms of improved accuracy and system resilience are not limited to selected cases: they also manifest as the scatterer's position changes.

\section{Conclusions and Future Work}
\label{sec:conclusions-and-future-work}
This paper establishes the critical role of frequency-dependent \ac{DMC}-aware modeling and processing in multi-band \ac{ISAC} for future 6G applications. It reveals that conventional intuitions about sensing drawn from \ac{SC}-only models do not necessarily hold, and that multi-band sensing performance gains in terms of localization accuracy and system resilience manifest in DMC-dominated scenarios. For future work, we plan to experimentally measure the \ac{DMC} with a multi-band \ac{ISAC} focus. We also plan to use a more physically accurate \ac{DMC} model so that parameters such as surface roughness and shape of extended targets can be estimated.

\section*{Acknowledgments}
\label{sec:acknowledgments}
This work is supported by Tamkeen under the Research Institute NYUAD grant CG017.

\bibliographystyle{IEEEtran}
\bibliography{refs}

\vspace{12pt}

\end{document}